\documentclass[12pt,a4paper]{article}

\usepackage{amsmath,a4,amssymb}
\usepackage[hypertex]{hyperref}

\makeatletter

\def\section{\@startsection{section}{1}{\z@}{-3.25ex plus -1ex minus
    -.2ex}{1.5ex plus .2ex}{\normalfont\bfseries}}
\def\subsection{\@startsection{subsection}{1}{\z@}{-3.25ex plus -1ex
    minus -.2ex}{1.5ex plus .2ex}{\normalfont\itshape}}

\makeatother

\allowdisplaybreaks[4]

\newcommand{\di}{\genfrac{}{}{0pt}{}}

\renewcommand{\title}[1]{\vspace{10mm}\noindent{\Large{\bf #1}}\vspace{8mm}}
\newcommand{\authors}[1]{\noindent{\large #1}\vspace{5mm}}
\newcommand{\address}[1]{{\itshape #1\vspace{2mm}}}

\begin{document}

\begin{titlepage}

\hspace*{\fill}hep-th/0402093

\begin{center}
  
\title{The $\beta$-function in duality-covariant 
noncommutative \\[1ex]$\phi^4$-theory}

\authors{Harald {\sc Grosse}$^1$ and Raimar {\sc Wulkenhaar}$^2$}

\address{$^{1}$\,Institut f\"ur Theoretische Physik, Universit\"at Wien\\
Boltzmanngasse 5, A-1090 Wien, Austria}

\address{$^{2}$\,Max-Planck-Institut f\"ur Mathematik in den
  Naturwissenschaften\\ 
Inselstra\ss{}e 22-26, D-04103 Leipzig, Germany}

\footnotetext[1]{harald.grosse@univie.ac.at}
\footnotetext[2]{raimar.wulkenhaar@mis.mpg.de}

\vskip 3cm

\textbf{Abstract}

\vskip 3mm

\begin{minipage}{14cm}%
  We compute the one-loop $\beta$-functions describing the
  renormalisation of the coupling constant $\lambda$ and the frequency
  parameter $\Omega$ for the real four-dimensional duality-covariant
  noncommutative $\phi^4$-model, which is renormalisable to all
  orders. The contribution from the one-loop four-point function is
  reduced by the one-loop wavefunction renormalisation, but the
  $\beta_\lambda$-function remains non-negative. Both $\beta_\lambda$
  and $\beta_\Omega$ vanish at the one-loop level for the
  duality-invariant model characterised by $\Omega=1$. Moreover,
  $\beta_\Omega$ also vanishes in the limit $\Omega\to 0$, which 
  defines the standard noncommutative $\phi^4$-quantum field theory. 
  Thus, the limit $\Omega \to 0$ exists at least at the one-loop
  level. 
\end{minipage}

\end{center}

\end{titlepage}

\section{Introduction}

For many years, the renormalisation of quantum field theories on
noncommutative $\mathbb{R}^4$ has been an open problem
\cite{Minwalla:1999px}. Recently, we have proven in
\cite{Grosse:2004yu} that the real duality-covariant $\phi^4$-model on
noncommutative $\mathbb{R}^4$ is renormalisable to all orders. The
duality transformation exchanges positions and momenta
\cite{Langmann:2002cc},
\begin{align}
\hat{\phi}(p) &\leftrightarrow \pi^2 \sqrt{|\det \theta|}\;\phi(x)\;, & 
p_\mu &\leftrightarrow \tilde{x}_\mu:= 2(\theta^{-1})_{\mu\nu} x^\nu\;,
\label{duality}
\end{align}
where $ \hat{\phi}(p_a)=\int d^4 x \;\mathrm{e}^{(-1)^a \mathrm{i}
  p_{a,\mu} x_a^\mu} \phi(x_a)$. The subscript $a$ refers to the
cyclic order in the $\star$-product. The duality-covariant
noncommutative $\phi^4$-action is given by
\begin{align}
S[\phi;\mu_0,\lambda,\Omega] := 
\int d^4x\;\Big( & \frac{1}{2} (\partial_\mu \phi) \star
(\partial^\mu \phi) + \frac{\Omega^2}{2} (\tilde{x}_\mu \phi) \star
(\tilde{x}^\mu \phi) + \frac{\mu_0^2}{2} \phi\star\phi
\nonumber
\\*
&+ \frac{\lambda}{4!} \phi \star \phi\star \phi \star \phi\Big)(x)\;.
\label{dual-action}
\end{align}
Under the transformation (\ref{duality}) one has 
\begin{align}
S\big[\phi;\mu_0,\lambda,\Omega\big] \mapsto 
\Omega^2
S\Big[\phi;\frac{\mu_0}{\Omega},\frac{\lambda}{\Omega^2},\frac{1}{\Omega}
\Big]\;. 
\label{dual-Omega}
\end{align}
In the special case $\Omega=1$ the action $S[\phi;\mu_0,\lambda,1]$ is
invariant under the duality (\ref{duality}). Moreover,
$S[\phi;\mu_0,\lambda,1]$ can be written as a standard matrix model
which is closely related to an exactly solvable model
\cite{Langmann:2003if}. 

Knowing that the action (\ref{dual-action}) gives rise to a
renormalisable quantum field theory \cite{Grosse:2004yu}, it is
interesting to compute the $\beta_\lambda$ and $\beta_\Omega$
functions which describe the renormalisation of the coupling constant
$\lambda$ and of the oscillator frequency $\Omega$. Whereas we have
proven the renormalisability in the Wilson-Polchinski approach
\cite{Wilson:1973jj, Polchinski:1983gv} adapted to non-local matrix
models \cite{Grosse:2003aj}, we compute the one-loop $\beta_\lambda$
and $\beta_\Omega$ functions by standard Feynman graph calculations.
Of course, these are Feynman graphs parametrised by matrix indices
instead of momenta. We rely heavily on the power-counting behaviour
proven in \cite{Grosse:2004yu}, which allows us to ignore in the
$\beta$-functions all non-planar graphs and the detailed index
dependence of the planar two- and four-point graphs.  Thus, only the
lowest-order (discrete) Taylor expansion of the planar two- and
four-point graphs can contribute to the $\beta$-functions. This means
that we cannot refer to the usual symmetry factors of commutative
$\phi^4$-theory so that we have to carefully recompute the graphs.

We obtain interesting consequences for the limiting cases $\Omega=1$
and $\Omega=0$ as discussed in Section~\ref{discussion}.

\section{Definition of the model}

The noncommutative $\mathbb{R}^4$ is defined as the algebra
$\mathbb{R}^4_\theta$ which as a vector space is given by the space
$\mathcal{S}(\mathbb{R}^4)$ of (complex-valued) Schwartz class functions of
rapid decay, equipped with the multiplication rule 
\begin{align}
  (a\star b)(x) &= \int \frac{d^4k}{(2\pi)^4} \int d^4 y \; a(x{+}\tfrac{1}{2}
  \theta {\cdot} k)\, b(x{+}y)\, \mathrm{e}^{\mathrm{i} k \cdot y}\;,
\label{starprod}
\\*
& (\theta {\cdot} k)^\mu = \theta^{\mu\nu} k_\nu\;,\quad k{\cdot}y = k_\mu
y^\mu\;,\quad \theta^{\mu\nu}=-\theta^{\nu\mu}\;.  \nonumber
\end{align}
We place ourselves into a coordinate system in which the only
non-vanishing components $\theta_{\mu\nu}$ are
$\theta_{12}=-\theta_{21}=\theta_{34}=-\theta_{42}=\theta$. 
We use an adapted base
\begin{align}
  b_{mn}(x) &= f_{m^1n^1}(x^1,x^2) \,f_{m^2n^2}(x^3,x^4) \;,\qquad
  m=\textstyle{\di{m^1}{m^2}}\in \mathbb{N}^2\,,~
n=\textstyle{\di{n^1}{n^2}} \in \mathbb{N}^2\;,
\label{bbas}
\end{align}
where the base $f_{m^1n^1}(x^1,x^2) \in \mathbb{R}^2_\theta$ is given in
\cite{Grosse:2003nw}. This base satisfies
\begin{align}
  (b_{mn} \star b_{kl})(x) &= \delta_{nk} b_{ml}(x) \;, & 
\int d^4x\, b_{mn}(x) =  4\pi^2 \theta^2 \delta_{mn} \;.
\end{align}
According to \cite{Grosse:2004yu}, the duality-covariant
$\phi^4$-action (\ref{dual-action}) expands as follows in the matrix
base (\ref{bbas}):
\begin{align}
  S[\phi;\mu_0,\lambda,\Omega] 
  &= 4\pi^2 \theta^2 \sum_{m,n,k,l \in \mathbb{N}^2} \Big(
  \frac{1}{2} G_{mn;kl} \phi_{mn} \phi_{kl} + \frac{\lambda}{4!}  \phi_{mn}
  \phi_{nk} \phi_{kl} \phi_{lm}\Big)\;,
\label{action}
\end{align}
where $\phi(x)=\sum_{m,n \in \mathbb{N}^2} \phi_{mn} b_{mn}(x)$ and 
\begin{align}
  G_{mn;kl} &= \Big(\mu_0^2{+} \frac{2}{\theta}(1{+}\Omega^2)
  (m^1{+}n^1{+}m^2{+}n^2{+}2) \Big)
  \delta_{n^1k^1} \delta_{m^1l^1} \delta_{n^2k^2} \delta_{m^2l^2} \nonumber
  \\*
  & - \frac{2}{\theta}(1{-}\Omega^2) \Big(\big(\sqrt{(n^1{+}1)(m^1{+}1)}\,
  \delta_{n^1+1,k^1}\delta_{m^1+1,l^1 } + \sqrt{n^1m^1}\, \delta_{n^1-1,k^1}
  \delta_{m^1-1,l^1} \big) \delta_{n^2k^2} \delta_{m^2l^2} \nonumber
  \\*
  & \qquad + \big(\sqrt{(n^2{+}1)(m^2{+}1)}\,
  \delta_{n^2+1,k^2}\delta_{m^2+1,l^2 } + \sqrt{n^2m^2}\, \delta_{n^2-1,k^2}
  \delta_{m^2-1,l^2} \big) \delta_{n^1k^1} \delta_{m^1l^1}\Big)\;.
\label{G4D}
\end{align}

The quantum field theory is defined by the partition function
\begin{align}
Z[J]&= \int \Big(\prod_{a,b \in \mathbb{N}^2} d
  \phi_{ab}\Big) \,\exp\big(-S[\phi]- 4\pi^2\theta^2 
\sum_{m,n \in \mathbb{N}^2} \phi_{mn} J_{nm}\big)\;.
\label{pathintm}
\end{align}
For the free theory defined by $\lambda=0$ in (\ref{action}), the
solution of (\ref{pathintm}) is given by
\begin{align}
Z[J]\big|_{\lambda=0} &= Z[0] \,\exp\Big( 4\pi^2 \theta^2 
\sum_{m,n,k,l \in  \mathbb{N}^2} \frac{1}{2} J_{mn} \Delta_{mn;kl}
J_{kl} \Big) \;,
\end{align}
where the propagator $\Delta$ is defined as the inverse of the kinetic
matrix $G$:
\begin{align}
  \sum_{k,l \in \mathbb{N}^2} G_{mn;kl} \Delta_{lk;sr} 
= \sum_{\in \mathbb{N}^2}
  \Delta_{nm;lk} G_{kl;rs} = \delta_{mr} \delta_{ns}\,.
\label{GD}
\end{align}
We have derived the propagator in \cite{Grosse:2004yu}:
\begin{align}
  &\Delta_{\di{m^1}{m^2}\di{n^1}{n^2}; \di{k^1}{k^2}\di{l^1}{l^2}} 
= \frac{\theta}{2(1{+}\Omega)^2} 
\delta_{m^1+k^1,n^1+l^1}\delta_{m^2+k^2,n^2+l^2}
  \nonumber
  \\*
  &\times \sum_{v^1=\frac{|m^1-l^1|}{2}}^{\frac{\min(m^1+l^1,n^1+k^1)}{2}}
  \sum_{v^2=\frac{|m^2-l^2|}{2}}^{\frac{\min(m^2+l^2,n^2+k^2)}{2}} B\big(1{+}
  \tfrac{\mu_0^2\theta}{8\Omega}
  {+}\tfrac{1}{2}(m^1{+}m^2{+}k^1{+}k^2){-}v^1{-}v^2, 1{+}2v^1{+}2v^2 \big)
  \nonumber
  \\*
  &\times {}_2F_1\bigg(\di{1{+} 2v^1{+}2v^2\,,\; \frac{\mu_0^2\theta}{8\Omega}
    {-}\frac{1}{2}(m^1{+}m^2{+}k^1{+}k^2){+}v^1{+}v^2 }{2{+}
    \frac{\mu_0^2\theta}{8\Omega}
    {+}\frac{1}{2}(m^1{+}m^2{+}k^1{+}k^2){+}v^1{+}v^2} \bigg|
  \frac{(1{-}\Omega)^2}{(1{+}\Omega)^2} \bigg) \nonumber
  \\*
  &\times \prod_{i=1}^2 \sqrt{ \binom{n^i}{v^i{+}\frac{n^i-k^i}{2}}
    \binom{k^i}{v^i{+}\frac{k^i-n^i}{2}} \binom{m^i}{v^i{+}\frac{m^i-l^i}{2}}
    \binom{l^i}{v^i{+}\frac{l^i-m^i}{2}}}
  \bigg(\frac{(1{-}\Omega)^2}{(1{+}\Omega)^2} \bigg)^{v^i} \,.
\label{noalpha}
\end{align}
Here, $B(a,b)$ is the Beta-function and
${}_2F_1\big(\di{a,b}{c}\big|z\big)$ the hypergeometric function. 

As usual we solve the interacting theory perturbatively:
\begin{align}
Z[J] &= Z[0] \exp\Big( - V\Big[\frac{\partial}{\partial J}\Big]\Big) 
\exp\Big( 4\pi^2 \theta^2 
\sum_{m,n,k,l \in  \mathbb{N}^2} \frac{1}{2} J_{mn} \Delta_{mn;kl}
J_{kl} \Big)\;,
\nonumber
\\*
V\Big[\frac{\partial}{\partial J}\Big] &:= 
\frac{\lambda}{4! (4\pi^2 \theta^2)^3}
\sum_{m,n,k,l \in \mathbb{N}^2} 
\frac{\partial^4}{\partial J_{ml} \,\partial J_{lk} \,
\partial J_{kn} \,\partial J_{nm}}\;.
\end{align}
It is convenient to pass to the generating functional of connected
Green's functions, $W[J]=\ln Z[J]$:
\begin{align}
W[J] &= \ln Z[0]+ W_{\mathrm{free}}[J] 
+ \ln \bigg(1+ \mathrm{e}^{-W_{\mathrm{free}}[J]}
\Big(\exp\Big( - V\Big[\frac{\partial}{\partial J}\Big]\Big) -1\Big)
\mathrm{e}^{W_{\mathrm{free}}[J]} \bigg)\;,
\nonumber
\\*
W_{\mathrm{free}}[J] &:= 
4\pi^2 \theta^2 
\sum_{m,n,k,l \in  \mathbb{N}^2} \frac{1}{2} J_{mn} \Delta_{mn;kl}
J_{kl} \;.
\end{align}
In order to obtain the expansion in $\lambda$ one has to expand $\ln
(1+x)$ as a power series in $x$ and $\exp(-V)$ as a power series in
$V$. By Legendre transformation we pass to the generating functional
of one-particle irreducible (1PI) Green's functions:
\begin{align}
\Gamma[\phi^{c\ell}] := 4\pi^2\theta^2 \sum_{m,n \in \mathbb{N}^2}
\phi^{c\ell}_{mn} J_{nm} - W[J] \;,
\end{align}
where $J$ has to be replaced by the inverse solution of 
\begin{align}
\phi^{c\ell}_{mn} := \frac{1}{4\pi^2\theta^2} 
\frac{\partial W[J]}{\partial J_{nm}}\;.
\label{phicl}
\end{align}

\section{Renormalisation group equation}

The computation of the expansion coefficients 
\begin{align}
\Gamma_{m_1n_1;\dots;m_Nn_N} := \frac{1}{N!} \frac{\partial^N 
\Gamma[\phi^{c\ell}]}{
\partial \phi^{c\ell}_{m_1n_1}\dots   
\partial \phi^{c\ell}_{m_Nn_N}} 
\label{Taylor}
\end{align}
of the effective action involves possibly divergent sums over 
undetermined loop indices. Therefore, we have to introduce a cut-off
$\mathcal{N}$ for all loop indices. According to \cite{Grosse:2004yu},
the expansion coefficients (\ref{Taylor}) can be decomposed into a
relevant/marginal and an irrelevant piece. As a result of the
renormalisation proof, the relevant/marginal parts have---after a
rescaling of the field amplitude---the same form as the initial action
(\ref{dual-action}), (\ref{action}) and (\ref{G4D}), now parametrised
by the ``physical'' mass, coupling constant and oscillator frequency:
\begin{align}
\Gamma_{\text{rel/marg}}[\mathcal{Z} \phi^{c\ell}] 
&= S\big[\phi^{c\ell}; \mu_{\text{phys}},
\lambda_{\text{phys}}, \Omega_{\text{phys}}\big]\;.
\label{GS}
\end{align}
In the renormalisation process, the physical quantities
$\mu^2_{\text{phys}}$, $\lambda_{\text{phys}}$ and
$\Omega_{\text{phys}}$ are kept constant with respect to the cut-off
$\mathcal{N}$. This is achieved by starting from a carefully adjusted
initial action
$S\big[\mathcal{Z}[\mathcal{N}]\phi,\mu_0[\mathcal{N}],\lambda[\mathcal{N}],
\Omega[\mathcal{N}]\Big]$, which gives rise to the bare effective
action $\Gamma[\phi^{c\ell};\mu_0[\mathcal{N}],
\lambda[\mathcal{N}],\Omega[\mathcal{N}],\mathcal{N}]$.
Expressing the bare parameters $\mu_0,\lambda,\Omega$ as a function of
the physical quantities and the cut-off, the expansion coefficients of
the renormalised effective action
\begin{align}
\Gamma^R [\phi^{c\ell};\mu_{\text{phys}},
\lambda_{\text{phys}},\Omega_{\text{phys}}]:= 
\Gamma\big[\mathcal{Z}[\mathcal{N}] \phi^{c\ell},\mu_0[\mathcal{N}],
\lambda[\mathcal{N}],\Omega[\mathcal{N}],\mathcal{N}\big]\Big|_{
\mu_{\text{phys}},
\lambda_{\text{phys}},\Omega_{\text{phys}}=\text{const}}
\end{align}
are finite and convergent in the limit $\mathcal{N}\to \infty$. In
other words,
\begin{align}
\lim_{\mathcal{N}\to \infty} \mathcal{N} \frac{d}{d \mathcal{N}}
  \Big(\mathcal{Z}^N[\mathcal{N}]
  \Gamma_{m_1n_1;\dots;m_Nn_N}[\mu_0[\mathcal{N}],
  \lambda[\mathcal{N}],\Omega[\mathcal{N}],\mathcal{N}]\Big)=0\;.
\end{align}
This implies the renormalisation group equation 
\begin{align}
\lim_{\mathcal{N}\to \infty} 
\Big(\mathcal{N} \frac{\partial}{\partial \mathcal{N}} 
+ N \gamma + \mu_0^2 \beta_{\mu_0} \frac{\partial}{\partial \mu_0^2} 
+ \beta_\lambda \frac{\partial}{\partial \lambda} 
+ \beta_\Omega \frac{\partial}{\partial \Omega} \Big) 
\Gamma_{m_1n_1;\dots;m_Nn_N}[\mu_0,\lambda,\Omega,\mathcal{N}] = 0\;,
\end{align}
where 
\begin{align}
\beta_{\mu_0} &=  \frac{1}{\mu_0^2}  \mathcal{N} 
\frac{\partial }{\partial \mathcal{N}} 
\Big( \mu_0^2[\mu_{\text{phys}},\lambda_{\text{phys}},
\Omega_{\text{phys}},\mathcal{N} ]\Big)\;,
\label{betaM}
\\
\beta_\lambda &=  \mathcal{N} 
\frac{\partial }{\partial \mathcal{N}}   
\Big( \lambda[\mu_{\text{phys}},\lambda_{\text{phys}},
\Omega_{\text{phys}},\mathcal{N} ]\Big)\;, 
\label{betaL}
\\
\beta_\Omega &= \mathcal{N} \frac{\partial }{\partial \mathcal{N}} 
\Big( \Omega[\mu_{\text{phys}},\lambda_{\text{phys}},
\Omega_{\text{phys}},\mathcal{N} ]\Big)\;,
\label{betaO}
\\
\gamma &= \mathcal{N} \frac{\partial }{\partial \mathcal{N}}
\Big( \ln \mathcal{Z}[\mu_{\text{phys}},\lambda_{\text{phys}},
\Omega_{\text{phys}},\mathcal{N} ]\Big)\;.
\label{gamma}
\end{align}

\section{One-loop computations}

Defining  $(\Delta J)_{mn}:=\sum_{p,q\in \mathbb{N}^2} \Delta_{mn;pq}
J_{pq}$ we write (parts of) the generating functional of connected
Green's functions up to second order in $\lambda$:
\begin{align}
W[J] &=\ln Z[0]+ 4\pi^2 \theta^2 
\sum_{m,n,k,l \in  \mathbb{N}^2} \frac{1}{2} J_{mn} \Delta_{mn;kl}
J_{kl} 
\nonumber
\\*
& - (4\pi^2\theta^2) \frac{\lambda}{4!} \sum_{m,n,k,l
  \in \mathbb{N}^2} \bigg\{
(\Delta J)_{ml} (\Delta J)_{lk} (\Delta J)_{kn} (\Delta J)_{nm} 
\nonumber
\\*
& \qquad\qquad+ \frac{1}{4\pi^2 \theta^2} \Big(  
\Delta_{nm;kn} (\Delta J)_{ml} (\Delta J)_{lk} 
+ \Delta_{kn;lk} (\Delta J)_{nm} (\Delta J)_{ml} 
\nonumber
\\*[-1ex]
&\qquad\qquad\qquad\qquad\qquad + \Delta_{nm;ml} (\Delta J)_{lk} 
(\Delta J)_{kn} 
+ \Delta_{lk;ml} (\Delta J)_{kn} (\Delta J)_{nm} \Big) 
\nonumber
\\*
& \qquad\qquad + \frac{1}{4\pi^2 \theta^2} \Big(  
\Delta_{nm;lk} (\Delta J)_{kn} (\Delta J)_{ml} 
+ \Delta_{kn;ml} (\Delta J)_{nm} (\Delta J)_{lk} \Big) 
\nonumber
\\*
& \qquad\qquad + \frac{1}{(4\pi^2 \theta^2)^2}
\Big( \big(\Delta_{nm;kn} \Delta_{lk;ml} 
+ \Delta_{kn;lk} \Delta_{nm;ml} \big)
+ \Delta_{nm;lk} \Delta_{kn;ml} \Big) \bigg\}
\nonumber
\\*
&+ \frac{\lambda^2}{2 (4!)^2} 
\sum_{m,n,k,l,r,s,t,u \in \mathbb{N}^2} \bigg\{\Big[\Big( 
\Delta_{ml;sr} \Delta_{lk;ts} (\Delta J)_{kn} (\Delta J)_{nm} 
+ \Delta_{ml;sr} \Delta_{kn;ts} (\Delta J)_{lk} (\Delta J)_{nm} 
\nonumber
\\*[-2.5ex]
& \hspace*{10em} 
+ \Delta_{ml;sr} \Delta_{nm;ts} (\Delta J)_{lk} (\Delta J)_{kn} 
+\Delta_{lk;sr} \Delta_{ml;ts} (\Delta J)_{kn} (\Delta J)_{nm} 
\nonumber
\\*
& \hspace*{10em} 
+ \Delta_{lk;sr} \Delta_{kn;ts} (\Delta J)_{ml} (\Delta J)_{nm} 
+ \Delta_{lk;sr} \Delta_{nm;ts} (\Delta J)_{ml} (\Delta J)_{kn} 
\nonumber
\\*
& \hspace*{10em} 
+ \Delta_{kn;sr} \Delta_{ml;ts} (\Delta J)_{lk} (\Delta J)_{nm} 
+ \Delta_{kn;sr} \Delta_{lk;ts} (\Delta J)_{ml} (\Delta J)_{nm} 
\nonumber
\\*
& \hspace*{10em} 
+ \Delta_{kn;sr} \Delta_{nm;ts} (\Delta J)_{ml} (\Delta J)_{lk} 
+ \Delta_{nm;sr} \Delta_{ml;ts} (\Delta J)_{lk} (\Delta J)_{kn} 
\nonumber
\\*
& \hspace*{10em} 
+ \Delta_{nm;sr} \Delta_{lk;ts} (\Delta J)_{ml} (\Delta J)_{kn} 
+ \Delta_{nm;sr} \Delta_{kn;ts} (\Delta J)_{ml} (\Delta J)_{lk} 
\Big)
\nonumber
\\*
& \hspace*{13em} 
\times (\Delta J)_{ru} (\Delta J)_{ut} 
\nonumber
\\*
& 
\hspace*{7em} + \text{5 permutations of $_{ts}$, 
$_{sr}$, $_{ru}$, $_{ut}$} \;\Big]
\nonumber
\\*
& \qquad+ \text{ 1PI-contributions with $\leq 2$ $J$'s }  
+ \text{ 1PR-contributions } \bigg\}
+ \mathcal{O}(\lambda^3)\;.
\end{align}
In second order in $\lambda$ we get a huge number of terms so that we 
display only the 1PI contribution with four $J$'s.

For the classical field (\ref{phicl}) we get 
$\phi^{c\ell}_{mn} = \sum_{p,q \in \mathbb{N}^2} \Delta_{nm;pq} J_{pq}  
+ \mathcal{O}(\lambda)$ so that 
\begin{align}
J_{pq} =  \sum_{r,s \in \mathbb{N}^2} G_{qp;rs} \phi^{c\ell}_{rs}
+ \mathcal{O}(\lambda)\;.
\label{Jpq}
\end{align}
The remaining part not displayed in (\ref{Jpq}) removes the
1PR-contributions when passing to $\Gamma[\phi^{c\ell}]$. 
We thus obtain
\begin{subequations}
\begin{align}
\Gamma[\phi^{c\ell}] &= \Gamma[0]
\nonumber
\\*
& + 4\pi^2\theta^2 \sum_{m,n,k,l \in \mathbb{N}^2} \frac{1}{2}
\Big\{G_{mn;kl}
+ \frac{\lambda}{6(4\pi^2\theta^2)} \Big(
\delta_{ml} \sum_{p \in \mathbb{N}^2} \Delta_{pn;kp} 
+ \delta_{kn} \sum_{p \in \mathbb{N}^2} \Delta_{mp;pl} \Big) 
\label{2P-plan}
\\*
&
\hspace*{10em} + \frac{\lambda}{6(4\pi^2\theta^2)} 
\Delta_{ml;kn} +\mathcal{O}(\lambda^2) 
\Big\} \phi^{c\ell}_{mn} \phi^{c\ell}_{kl} 
\label{2P-nonplan}
\\
& +   4\pi^2\theta^2 
\sum_{m,n,k,l,r,s,t,u \in \mathbb{N}^2} \frac{\lambda}{4!} 
\Big\{\delta_{nk} \delta_{lr} \delta_{st} \delta_{um} 
\label{4P-plan1}
\\*
& - \frac{\lambda}{2 (4!) (4\pi^2 \theta^2)}
\Big( \sum_{p,q \in \mathbb{N}^2} \big(
4 \Delta_{mp;qs} \Delta_{pl;tq} \delta_{kn} \delta_{ur} 
+ 4 \Delta_{kp;qs} \Delta_{pn;tq} \delta_{ml} \delta_{ur} 
\nonumber
\\*[-2ex]
& \hspace*{10em} 
+ 4 \Delta_{pl;rq} \Delta_{mp;qu} \delta_{nk} \delta_{st}
+ 4 \Delta_{pn;rq} \Delta_{kp;qu} \delta_{ml} \delta_{st}\big)
\label{4P-plan2}
\\*[0.5ex]
& \hspace*{6.5em} 
+ \sum_{p \in \mathbb{N}^2} \big( 4 \Delta_{ml;ps} \Delta_{kn;tp} \delta_{ur} 
+ 4 \Delta_{kn;ps} \Delta_{ml;tp} \delta_{ur} 
+ 4 \Delta_{mp;ts} \Delta_{pl;ru} \delta_{nk} 
\nonumber
\\*[-2ex]
& \hspace*{10em} 
+ 4 \Delta_{pl;ts} \Delta_{mp;ru} \delta_{nk} 
+ 4 \Delta_{kp;ts} \Delta_{pn;ru} \delta_{ml} 
+ 4 \Delta_{pn;ts} \Delta_{kp;ru} \delta_{ml} 
\nonumber
\\*
& \hspace*{10em} 
+ 4 \Delta_{ml;rp} \Delta_{kn;pu} \delta_{st}
+ 4 \Delta_{kn;rp} \Delta_{ml;pu} \delta_{st}\big)
\label{4P-nonplan1}
\\*[0.5ex]
& \hspace*{6.5em} 
+ \sum_{p,q \in \mathbb{N}^2} \big( 
4\Delta_{pl;qs} \Delta_{mp;tq} \delta_{nk} \delta_{ur} 
+ 4\Delta_{pn;qs} \Delta_{kp;tq} \delta_{ml} \delta_{ur} 
\nonumber
\\*[-2ex]
& \hspace*{10em} 
+ 4 \Delta_{kp;rq} \Delta_{pn;qu} \delta_{ml} \delta_{st}
+ 4 \Delta_{mp;rq} \Delta_{pl;qu} \delta_{nk} \delta_{st} \big)
\label{4P-nonplan2}
\\*
& \hspace*{6.5em} 
+ 4 \Delta_{ml;ts} \Delta_{kn;ru} 
+ 4 \Delta_{kn;ts} \Delta_{ml;ru} 
\Big) + \mathcal{O}(\lambda^2)
\Big\}\phi^{c\ell}_{mn} \phi^{c\ell}_{kl}
\phi^{c\ell}_{rs} \phi^{c\ell}_{tu} 
\label{4P-nonplan3}
\\*
& + \mathcal{O}(\lambda^2)\;.
\nonumber
\end{align}
\end{subequations}
Here, (\ref{2P-plan}) contains the contribution to the planar
two-point function and (\ref{2P-nonplan}) the contribution to the
non-planar two-point function. Next, (\ref{4P-plan1}) and
(\ref{4P-plan2}) contribute to the planar four-point function, whereas
(\ref{4P-nonplan1}), (\ref{4P-nonplan2}) and (\ref{4P-nonplan3})
constitute three different types of non-planar four-point functions.

Introducing the cut-off $p^i,q^i \leq \mathcal{N}$ in the internal
sums over $p,q \in \mathbb{N}^2$, we split the effective action
according to \cite{Grosse:2004yu} as follows into a relevant/marginal
and an irrelevant piece ($\Gamma[0]$ can be ignored):
\begin{align}
\Gamma[\phi^{c\ell}] & \equiv   
\Gamma_{\text{rel/marg}}[\phi^{c\ell}] + 
\Gamma_{\text{irrel}}[\phi^{c\ell}] \;,
\\
\Gamma_{\text{rel/marg}}[\phi^{c\ell}] &= 
 4\pi^2\theta^2 \!\!\! \sum_{m,n,k,l \in \mathbb{N}^2} \frac{1}{2}
\Big\{G_{mn;kl}
+ \frac{\lambda}{6(4\pi^2\theta^2)} \delta_{ml}\delta_{kn} 
\Big(
2 \sum_{p^1,p^2=0}^N \Delta_{\di{0}{0}\di{p^1}{p^2};\di{p^1}{p^2}\di{0}{0}} 
\nonumber
\\*
&\qquad + (m^1{+}n^1{+}m^2{+}n^2) \sum_{p^1,p^2=0}^N \big(
\Delta_{\di{1}{0}\di{p^1}{p^2};\di{p^1}{p^2}\di{1}{0}} 
-\Delta_{\di{0}{0}\di{p^1}{p^2};\di{p^1}{p^2}\di{0}{0}} \big)
\Big) +\mathcal{O}(\lambda^2) 
\Big\} \phi^{c\ell}_{mn} \phi^{c\ell}_{kl} 
\nonumber
\\*
& +   4\pi^2\theta^2 \!\!\!\!
\sum_{m,n,k,l \in \mathbb{N}^2} \frac{\lambda}{4!} 
\Big\{1 - \frac{\lambda}{3 (4\pi^2 \theta^2)}
\sum_{p^1,p^2=0}^N \big(
\Delta_{\di{0}{0}\di{p^1}{p^2};\di{p^1}{p^2}\di{0}{0}} \big)^2 
+\mathcal{O}(\lambda^2)\Big\} \phi^{c\ell}_{mn}
\phi^{c\ell}_{nk} \phi^{c\ell}_{kl} \phi^{c\ell}_{lm}\;.
\label{Gammarel}
\end{align}
To the marginal four-point function and the relevant two-point
function there contribute only the projections to planar graphs with
vanishing external indices.  The marginal two-point function is given
by the next-to-leading term in the discrete Taylor expansion around
vanishing external indices.

In a regime where $\lambda[\mathcal{N}]$ is so small that the perturbative
expansion is valid in (\ref{Gammarel}), the irrelevant part
$\Gamma_{\text{irrel}}$ can be completely ignored. Comparing
(\ref{Gammarel}) with the initial action according to 
(\ref{dual-action}),(\ref{action}) and (\ref{G4D}), we have 
$\Gamma_{\text{rel/marg}}[\mathcal{Z} \phi^{c\ell}] 
= S\big[\phi^{c\ell}; \mu_{\text{phys}},
\lambda_{\text{phys}}, \Omega_{\text{phys}}\big]$ with 
\begin{align}
\mathcal{Z} &= 1-\frac{\lambda}{192 \pi^2 \theta}  
\sum_{p^1,p^2=0}^{\mathcal{N}} \big(
\Delta_{\di{1}{0}\di{p^1}{p^2};\di{p^1}{p^2}\di{1}{0}} 
-\Delta_{\di{0}{0}\di{p^1}{p^2};\di{p^1}{p^2}\di{0}{0}} \big)
+ \mathcal{O}(\lambda^2)\;, 
\label{ZN}
\\
\mu^2_{\text{phys}} &= \mu_0^2 \Big(1 + \frac{\lambda}{
12 \pi^2 \theta^2\mu_0^2} \sum_{p^1,p^2=0}^{\mathcal{N}} \big(
2\Delta_{\di{0}{0}\di{p^1}{p^2};\di{p^1}{p^2}\di{0}{0}} 
-\Delta_{\di{1}{0}\di{p^1}{p^2};\di{p^1}{p^2}\di{1}{0}} \big)
\nonumber
\\*
&\qquad\qquad 
- \frac{\lambda}{96 \pi^2  \theta} \sum_{p^1,p^2=0}^{\mathcal{N}} \big(
\Delta_{\di{1}{0}\di{p^1}{p^2};\di{p^1}{p^2}\di{1}{0}} 
-\Delta_{\di{0}{0}\di{p^1}{p^2};\di{p^1}{p^2}\di{0}{0}} \big)
+ \mathcal{O}(\lambda^2)\Big)\;,
\label{mph}
\\
\lambda_{\text{phys}} &= \lambda \Big(1 
-  \frac{\lambda}{12\pi^2 \theta^2}\sum_{p^1,p^2=0}^{\mathcal{N}} \big(
\Delta_{\di{0}{0}\di{p^1}{p^2};\di{p^1}{p^2}\di{0}{0}} \big)^2 
\nonumber
\\
& \qquad\qquad 
- \frac{\lambda}{48 \pi^2 \theta} 
\sum_{p^1,p^2=0}^{\mathcal{N}} \big(
\Delta_{\di{1}{0}\di{p^1}{p^2};\di{p^1}{p^2}\di{1}{0}} 
-\Delta_{\di{0}{0}\di{p^1}{p^2};\di{p^1}{p^2}\di{0}{0}} \big)
+ \mathcal{O}(\lambda^2)\Big)\;, 
\label{lph}
\\
\Omega_{\text{phys}} &= \Omega\Big(1 + \frac{\lambda(1{-}\Omega^2)}{192 \pi^2
  \theta \Omega^2} \sum_{p^1,p^2=0}^{\mathcal{N}} \big(
\Delta_{\di{1}{0}\di{p^1}{p^2};\di{p^1}{p^2}\di{1}{0}} 
-\Delta_{\di{0}{0}\di{p^1}{p^2};\di{p^1}{p^2}\di{0}{0}} \big)
+ \mathcal{O}(\lambda^2)\Big)\;. 
\label{Oph}
\end{align}

Solving (\ref{mph}), (\ref{lph}) and (\ref{Oph}) for the
bare quantities, we obtain to one-loop order
\begin{align}
&\mu_0^2[\mu_{\text{phys}},\lambda_{\text{phys}},\Omega_{\text{phys}},
\mathcal{N} ] 
\nonumber
\\*
&\qquad = \mu_{\text{phys}}^2 \Big(1 - 
 \frac{\lambda_{\text{phys}}}{12\pi^2 \theta^2 \mu_{\text{phys}}^2} 
\sum_{p^1,p^2=0}^{\mathcal{N}}
\Delta_{\di{0}{0}\di{p^1}{p^2};\di{p^1}{p^2}\di{0}{0}} 
\nonumber
\\*[-1ex]
& \qquad\qquad\qquad
+ \frac{\lambda_{\text{phys}}}{96\pi^2 \theta} 
\Big(1+\frac{8}{\theta \mu_{\text{phys}}^2}\Big)
\sum_{p^1,p^2=0}^{\mathcal{N}} \big(
\Delta_{\di{1}{0}\di{p^1}{p^2};\di{p^1}{p^2}\di{1}{0}} 
-\Delta_{\di{0}{0}\di{p^1}{p^2};\di{p^1}{p^2}\di{0}{0}} \big)
+ \mathcal{O}(\lambda_{\text{phys}}^2)\Big)\;,
\\[0.5ex]
&\lambda[\mu_{\text{phys}},\lambda_{\text{phys}},\Omega_{\text{phys}},
\mathcal{N} ] 
\nonumber
\\*
&\qquad = \lambda_{\text{phys}} \Big(1 
+  \frac{\lambda_{\text{phys}}}{12\pi^2 \theta^2}
\sum_{p^1,p^2=0}^{\mathcal{N}} \big(
\Delta_{\di{0}{0}\di{p^1}{p^2};\di{p^1}{p^2}\di{0}{0}} \big)^2 
+ \frac{\lambda_{\text{phys}}}{48 \pi^2 \theta} 
\sum_{p^1,p^2=0}^{\mathcal{N}} \big(
\Delta_{\di{1}{0}\di{p^1}{p^2};\di{p^1}{p^2}\di{1}{0}} 
-\Delta_{\di{0}{0}\di{p^1}{p^2};\di{p^1}{p^2}\di{0}{0}} \big)
\nonumber
\\*[-2ex]
& \qquad
+ \mathcal{O}(\lambda_{\text{phys}}^2)\Big)\;,
\label{lambdaN}
\\[0.5ex]
&\Omega[\mu_{\text{phys}}, \lambda_{\text{phys}},\Omega_{\text{phys}},
\mathcal{N} ] 
\nonumber
\\*
&\qquad = \Omega_{\text{phys}} \Big(1 
 - \frac{\lambda_{\text{phys}}(1{-}\Omega_{\text{phys}}^2)}{192 \pi^2
  \theta \Omega^2_{\text{phys}}} \sum_{p^1,p^2=0}^{\mathcal{N}} \big(
\Delta_{\di{1}{0}\di{p^1}{p^2};\di{p^1}{p^2}\di{1}{0}} 
-\Delta_{\di{0}{0}\di{p^1}{p^2};\di{p^1}{p^2}\di{0}{0}} \big)
+ \mathcal{O}(\lambda_{\text{phys}}^2)\Big)\;.
\end{align}

Inserting (\ref{noalpha}) into (\ref{lambdaN}) we can now compute the
$\beta_\lambda$-function (\ref{betaL}) up to one-loop order, 
omitting the index $_{\text{phys}}$ on $\mu^2$ and $\Omega$ for
simplicity: 
\begin{align}
\beta_\lambda &= 
\frac{\lambda_{\text{phys}}^2}{48 \pi^2} 
\mathcal{N} \frac{\partial}{\partial \mathcal{N}} 
\sum_{p^1,p^2=0}^{\mathcal{N}} \Bigg\{
\left(\raisebox{-1ex}{\mbox{$\dfrac{
{}_2F_1\Big(\di{1\,,\; \frac{\mu_0^2\theta}{8\Omega}
-\frac{1}{2}(p^1{+}p^2)}{2+\frac{\mu_0^2\theta}{8\Omega}
+\frac{1}{2}(p^1{+}p^2)}\Big| \frac{(1{-}\Omega)^2}{(1{+}\Omega)^2}\Big)
}{(1{+}\Omega)^2(1+\frac{\mu_0^2\theta}{8\Omega}
+\frac{1}{2}(p^1{+}p^2))}$}} \right)^{\!\!2}
\nonumber
\\*
& \qquad + \dfrac{p^1(1{-}\Omega)^2 \;
{}_2F_1\Big(\di{3\,,\; \frac{1+\mu_0^2\theta}{8\Omega}
-\frac{1}{2}(p^1{+}p^2{+}1)}{3+\frac{\mu_0^2\theta}{8\Omega}
+\frac{1}{2}(p^1{+}p^2{+}1)}\Big| \frac{(1{-}\Omega)^2}{(1{+}\Omega)^2}\Big)
}{(1{+}\Omega)^4
\big(\frac{1}{2}{+}\frac{\mu_0^2\theta}{8\Omega}{+}\frac{1}{2}(p^1{+}p^2)\big)
\big(\frac{3}{2}{+}\frac{\mu_0^2\theta}{8\Omega}{+}\frac{1}{2}(p^1{+}p^2)\big)
\big(\frac{5}{2}{+}\frac{\mu_0^2\theta}{8\Omega}{+}\frac{1}{2}(p^1{+}p^2)\big)
}
\nonumber
\\
&\qquad + \dfrac{
{}_2F_1\Big(\di{1\,,\; \frac{\mu_0^2\theta}{8\Omega}
-\frac{1}{2}(p^1{+}p^2{+}1)}{2+\frac{\mu_0^2\theta}{8\Omega}
+\frac{1}{2}(p^1{+}p^2{+}1)}\Big| \frac{(1{-}\Omega)^2}{(1{+}\Omega)^2}\Big)
}{2(1{+}\Omega)^2(\frac{3}{2}+\frac{\mu_0^2\theta}{8\Omega}
+\frac{1}{2}(p^1{+}p^2))}
-\dfrac{
{}_2F_1\Big(\di{1\,,\; \frac{\mu_0^2\theta}{8\Omega}
-\frac{1}{2}(p^1{+}p^2)}{2+\frac{\mu_0^2\theta}{8\Omega}
+\frac{1}{2}(p^1{+}p^2)}\Big| \frac{(1{-}\Omega)^2}{(1{+}\Omega)^2}\Big)
}{2(1{+}\Omega)^2 (1+\frac{\mu_0^2\theta}{8\Omega}
+\frac{1}{2}(p^1{+}p^2))}
+ \mathcal{O}(\lambda_{\text{phys}})\Big) \Bigg\}\;.
\label{betaF}
\end{align}
Symmetrising the numerator in the second line $p^1\mapsto
\frac{1}{2}(p^1{+}p^2)$ and using the expansions 
\begin{align}
{}_2F_1\Big(\di{1\,,\;a-p}{b+p}\Big|z\Big) &= \frac{1}{1{+}z} +
\frac{z(a{+}b)+z^2(a{+}b{-}2)}{p(1{+}z)^3} +
\mathcal{O}(p^{-2})\;, 
\nonumber
\\*
{}_2F_1\Big(\di{3\,,\;a-p}{b+p}\Big|z\Big) &= \frac{1}{(1{+}z)^3} +
\mathcal{O}(p^{-1})\;, 
\label{expF}
\end{align}
which are valid for large $p$, we obtain up to irrelevant contributions
vanishing in the limit $\mathcal{N}\to\infty$ 
\begin{align}
\beta_\lambda &= 
\frac{\lambda_{\text{phys}}^2}{48 \pi^2} 
 \mathcal{N} \frac{\partial}{\partial \mathcal{N}} 
\sum_{p^1,p^2=0}^{\mathcal{N}} 
\frac{1}{(1{+}\Omega_{\text{phys}}^2)^2}
\frac{1}{\big(1+p^1{+}p^2\big)^2}
\Big\{1 + \frac{(1{-}\Omega_{\text{phys}}^2)^2}{2(1{+}\Omega_{\text{phys}}^2)}
-\frac{(1{+}\Omega_{\text{phys}}^2)}{2} \Big\}
\nonumber
\\*[-0.5ex] 
& + \mathcal{O}(\lambda_{\text{phys}}^3) 
+ \mathcal{O}(\mathcal{N}^{-1})\nonumber
\\*[0.5ex]
&=  \frac{\lambda_{\text{phys}}^2}{48 \pi^2} 
\frac{(1{-}\Omega_{\text{phys}}^2)}{(1{+}\Omega_{\text{phys}}^2)^3}
+ \mathcal{O}(\lambda_{\text{phys}}^3)+ \mathcal{O}(\mathcal{N}^{-1}) \;.
\end{align}
Similarly, one obtains
\begin{align}
\beta_\Omega  &=  \frac{\lambda_{\text{phys}} \Omega_{\text{phys}}}{96 \pi^2} 
\frac{(1{-}\Omega_{\text{phys}}^2)}{(1{+}\Omega_{\text{phys}}^2)^3}
+ \mathcal{O}(\lambda_{\text{phys}}^2) + \mathcal{O}(\mathcal{N}^{-1}) \;,
\\
\beta_{\mu_0}
&=   - \frac{\lambda_{\text{phys}}}{48 \pi^2 \theta
  \mu_{\text{phys}}^2 (1{+}\Omega_{\text{phys}}^2) }
\Big(4\mathcal{N}\ln(2)
+ \frac{(8{+} \theta\mu_{\text{phys}}^2)\Omega^2_{\text{phys}}}{
(1{+}\Omega_{\text{phys}}^2)^2} \Big)
\nonumber
\\*
&+ \mathcal{O}(\lambda_{\text{phys}}^2)
+ \mathcal{O}(\mathcal{N}^{-1})\;,
\label{betamu}
\\
\gamma  &=  \frac{\lambda_{\text{phys}} }{96 \pi^2} 
\frac{\Omega^2_{\text{phys}}}{(1{+}\Omega_{\text{phys}}^2)^3}
+ \mathcal{O}(\lambda_{\text{phys}}^2) + \mathcal{O}(\mathcal{N}^{-1}) \;.
\end{align}

\section{Discussion}
\label{discussion}

We have computed the one-loop $\beta$- and $\gamma$-functions in real
four-dimensional duality-covariant noncommutative $\phi^4$-theory.
Remarkably, this model has a one-loop contribution to the wavefunction
renormalisation which compensates partly the contribution from the
planar one-loop four-point function to the $\beta_\lambda$-function.
The one-loop $\beta_\lambda$-function is non-negative and vanishes in
the distinguished case $\Omega=1$ of the duality-invariant model, see
(\ref{dual-Omega}). At $\Omega=1$ also the $\beta_\Omega$-function
vanishes. This is of course expected (to all orders), because for
$\Omega=1$ the propagator (\ref{noalpha}) is diagonal,
$\Delta_{\di{m^1}{m^2}\di{n^1}{n^2}; \di{k^1}{k^2}\di{l^1}{l^2}}
\big|_{\Omega=1}= \frac{
  \delta_{m^1l^1}\delta_{k^1n^1}\delta_{m^2l^2}\delta_{k^2n^2}}{
  \mu_0^2 + (4/\theta) (m^1{+}m^2{+}n^1{+}n^2{+}2)}$, so that the
Feynman graphs never generate terms with $|m^i-l^i|=|n^i-k^i|=1$ in
(\ref{G4D}).

The similarity of the duality-invariant theory with the exactly
solvable models discussed in \cite{Langmann:2003if} suggests that also
the $\beta_\lambda$-function vanishes to all orders for $\Omega=1$.
The crucial differences between our model with $\Omega=1$ and
\cite{Langmann:2003if} is that we are using \emph{real} fields, for
which it is not so clear that the construction of
\cite{Langmann:2003if} can be applied. But the planar graphs of a real
and a complex $\phi^4$-model are very similar so that we expect
identical $\beta_\lambda$-functions (possibly up to a global factor)
for the complex and the real model. Since a main feature of
\cite{Langmann:2003if} was the independence on the dimension of the
space, the model with $\Omega=1$ and matrix cut-off $\mathcal{N}$
should be (more or less) equivalent to a two-dimensional model, which
has a mass renormalisation only \cite{Grosse:2003nw}. Therefore, we
conjecture a vanishing $\beta_\lambda$-function in four-dimensional
duality-invariant noncommutative $\phi^4$-theory to all orders.

The most surprising result is that the one-loop
$\beta_\Omega$-function also vanishes for $\Omega \to 0$. We cannot
directly set $\Omega=0$, because the hypergeometric functions in
(\ref{betaF}) become singular and the expansions (\ref{expF}) are not
valid. Moreover, the power-counting theorems of \cite{Grosse:2004yu},
which we used to project to the relevant/marginal part of the
effective action (\ref{Gammarel}), also require $\Omega>0$. However,
in the same way as in the renormalisation of two-dimensional
noncommutative $\phi^4$-theory \cite{Grosse:2003nw}, it is possible to
switch off $\Omega$ very weakly with the cut-off $\mathcal{N}$, e.g.\ with
\begin{align}
\Omega=\mathrm{e}^{-\big(\ln(1+ \ln (1+\mathcal{N}))\big)^2}\;.   
\label{Omega-off}
\end{align}
The decay (\ref{Omega-off}) for large $\mathcal{N}$ over-compensates
the growth of any polynomial in $\ln \mathcal{N}$, which according to
\cite{Grosse:2004yu} is the bound for the graphs contributing to a
renormalisation of $\Omega$. On the other hand, (\ref{Omega-off}) does
not modify the expansions (\ref{expF}). Thus, in the limit $\mathcal{N} \to
\infty$, we have constructed the usual noncommutative $\phi^4$-theory
given by $\Omega=0$ in (\ref{dual-action}) at the one-loop level. It
would be very interesting to know whether this construction of the
noncommutative $\phi^4$-theory as the limit of a sequence
(\ref{Omega-off}) of duality-covariant $\phi^4$-models can be extended
to higher loop order.

We also notice that the one-loop $\beta_\lambda$- and
$\beta_\Omega$-functions are independent of the noncommutativity scale
$\theta$. There is, however a contribution to the one-loop mass
renormalisation via the dimensionless quantity $\mu_{\text{phys}}^2
\theta$, see (\ref{betamu}).

\section*{Acknowledgement}

We thank Helmut Neufeld for interesting discussions about the calculation
of $\beta$-functions.


\begin{thebibliography}{9}

\bibitem{Minwalla:1999px}
S.~Minwalla, M.~Van Raamsdonk and N.~Seiberg,
``Noncommutative perturbative dynamics,''
JHEP {\bf 0002} (2000) 020
[arXiv:hep-th/9912072].

\bibitem{Grosse:2004yu} H.~Grosse and R.~Wulkenhaar, ``Renormalisation
  of $\phi^4$ theory on noncommutative $\mathbb{R}^4$ in the matrix
  base,'' arXiv:hep-th/0401128.

\bibitem{Langmann:2002cc}
E.~Langmann and R.~J.~Szabo,
``Duality in scalar field theory on noncommutative phase spaces,''
Phys.\ Lett.\ B {\bf 533} (2002) 168
[arXiv:hep-th/0202039].

\bibitem{Langmann:2003if}
E.~Langmann, R.~J.~Szabo and K.~Zarembo,
``Exact solution of quantum field theory on noncommutative phase spaces,''
JHEP {\bf 0401} (2004) 017
[arXiv:hep-th/0308043].

\bibitem{Wilson:1973jj}
K.~G.~Wilson and J.~B.~Kogut,
``The Renormalization Group And The Epsilon Expansion,''
Phys.\ Rept.\  {\bf 12} (1974) 75.

\bibitem{Polchinski:1983gv}
J.~Polchinski,
``Renormalization And Effective Lagrangians,''
Nucl.\ Phys.\ B {\bf 231} (1984) 269.

\bibitem{Grosse:2003aj}
H.~Grosse and R.~Wulkenhaar,
``Power-counting theorem for non-local matrix models and renormalisation,''
arXiv:hep-th/0305066.

\bibitem{Grosse:2003nw} H.~Grosse and R.~Wulkenhaar, ``Renormalisation
  of $\phi^4$ theory on noncommutative $\mathbb{R}^2$ in the matrix
  base,'' JHEP {\bf 0312} (2003) 019 
[arXiv:hep-th/0307017].

\end{thebibliography}
\end{document}